# Safety Challenges and Analysis of Autonomous Electric Vehicle Development: Insights from On-Road Testing and Accident Reports

*Qasim Ajao
Georgia Southern University
Department of Electrical Engineering
Statesboro, Georgia USA

Olukotun Oludamilare
Georgia Southern University
Deapartment of Electrical Engineering
Statesboro, Georgia USA

**Abstract**: Autonomous electric vehicles (AEVs) hold great promise for the future of automotive engineering, but safety remains a significant challenge in their development and commercialization. Therefore, conducting a comprehensive analysis of AEV development and reported accidents is crucial. This paper reviews the levels of automation in AEVs, their disengagement frequencies, and on-road accident reports. According to the report, numerous manufacturers thoroughly tested AEVs across a distance of more than 3.9 million miles between 2014 and 2022. Disengagement frequencies vary among manufacturers, and approximately 65% of accidents during this period occurred while AEVs were operating in autonomous mode. Notably, the majority of accidents (90%) were caused by other road users, with only a small fraction (~8%) directly attributed to AEVs. Enhancing AEVs' ability to detect and mitigate safety risks from external sources has the potential to significantly improve their safety. This paper provides valuable insights into AEV safety by emphasizing the importance of comprehensively understanding AEV development and reported accidents. Through the analysis of disengagement and accident reports, the study highlights the prevalence of passive accidents caused by other road users. Future research should concentrate on enabling AEVs to effectively detect and respond to safety risks originating from external sources to enhance AEV safety. Overall, this analysis contributes to the ongoing efforts in AEV development and provides guidance for strategies aimed at improving their safety features.

## 1. INTRODUCTION

Autonomous driving technology has come to light as a possible option as society places more emphasis on minimizing traffic accidents, congestion, energy consumption, and pollutants. Autonomous electric vehicles (AEVs) equipped with advanced technologies can assist or operate independently, reducing the need for human intervention in vehicle control. The level of automation in AEVs determines whether control decisions are made by a human driver or an autonomous system based on the vehicle's capabilities and the surrounding environment. Implementing autonomous technology in transportation systems offers significant opportunities to address economic and environmental challenges. AEVs can enhance road safety by minimizing human errors that contribute to the majority of accidents. They also improve commuting by allowing occupants to engage in other activities and optimizing traffic paths and parking. AEVs promote mobility for individuals with disabilities, reduce the burden on mass transit, alleviate congestion, save fuel through efficient fleet management, and reduce stress for commuters. Additionally, they have the potential to save energy, decrease emissions, and positively impact pavement sustainability by minimizing crashes and optimizing vehicle operation [1].

The development of AEVs has been driven by the potential benefits of autonomous technology. Research has evolved from infrastructure-centered to vehicle-centered approaches, involving private companies like Google, Audi, Toyota, and Nissan. Road testing of AEV technologies has gained momentum, with features like lane-keeping, collision avoidance, and adaptive cruise control already implemented. However, safety concerns prevent the full commercialization of fully autonomous vehicles. Optimism about AEV safety varies across demographics and countries, and addressing safety risks and human factors in vehicle-human interaction is crucial. Regulations must adapt to technological progress [2]. A thorough understanding of automation levels, incidents, and on-road testing status is required to progress AEV technology. Conducting a thorough investigation into AEV-related accidents and predicting potential accidents as AEV technology advances is crucial. While significant efforts have been made in AEV technology development, a comprehensive statistical analysis of safety issues is lacking. Understanding system failures and causes through critical analysis is essential for AEV design and development. This study aims to systematically analyze safety issues in autonomous technology for vehicles, providing valuable insights to stakeholders and advancing AEV technology.

## 2. DEGREE OF AUTOMATION

To minimize the impact of autonomous electric vehicles (AEVs) on traditional road users, such as vehicles, pedestrians, bicyclists, and construction workers, regulators need to establish a clear definition of AEVs. As mentioned earlier, the level of automation in AEVs is determined by factors like the complexity of the autonomous technology used, the range of environmental perception, and the involvement of human drivers or vehicle systems in making driving decisions. These factors directly impact the safety of AEVs [3]. This section provides a summary and comparison of different organizations' definitions of automation levels. The classic concept of automation levels specifies ten stages of automation based on the responsibilities of human operators and vehicle systems in the driving process. It was first put forth by Sheridan and Verplank in 1987 and then amended by Parasuraman et al. in 2000. Level 1 denotes complete human decision and action-making with no automation. Alternate decisions or action plans may be suggested by the system in Levels 2 to 4, but human supervisors must determine whether to follow them or not. As of Level 5, the system can carry out decisions with a human operator's consent [1].

Level 6 offers the human driver a restricted period of time to respond before taking autonomous action. Level 7 alerts the human supervisor following an automatic action, but Level 8 only provides information upon request. Level 9 is concerned with the system determining whether to notify a human supervisor following an automatic action, whereas Level 10 involves complete automation that ignores human variables. There are further resources where you may learn more about these ten levels of automation. The aviation engineering framework Pilot Authorization and Control of Tasks (PACT) includes six degrees of automation. According to this hypothesis, systems at Level 5 can run completely autonomously but can still be overridden by a human pilot because Level 0, which denotes no computer autonomy, is still feasible. Furthermore, depending on how human pilots and technologies interact operationally, the PACT framework proposes four aided modes. More information on these six levels can be found in the mentioned source [1, 4].

The National Highway Traffic Safety Administration (NHTSA) of the United States has set up a hierarchical framework with five levels to classify automation in the field of car engineering. The numbers from 0 to 4 are used to name these stages. Level 0 cars don't have any kind of technology, so the driver is in charge of everything. Level 4 is the highest level of automation [5]. This is where self-driving cars can watch their surroundings and do all the important driving tasks on their own. Level 3 means that the car can drive itself some of the time, but the driver can still take control in certain situations. Most ongoing projects to build self-driving cars are in line with Level 3. The Society of Automotive Engineers (SAE) is the most trusted source for widely used meanings of terms related to automated driving. The SAE standards have been taken on by NHTSA, and they are regularly updated. SAE divides the levels of automation in cars into six groups based on how much human participation is needed by the automation system. These levels range from 0 (no automation) to 5 (full automation), with 0 being no automation and 5 being full automation. These rules are often used by regulators, lawmakers, and automakers in their work.

The different levels of automation are based on how the automation system and human drivers work together to handle steering, throttle control, monitoring the environment, falling back to dynamic driving tasks (DDT), and the system's ability to switch between different autonomous driving modes. Levels 0 to 2 depend on human workers to do some or all of the dynamic driving tasks (DDT), while Levels 3 to 5 show conditional automation, high automation, and full automation, respectively. These higher levels show that the system can handle all dynamic driving tasks (DDT) on its own while it is in action. The Society of Automotive Engineers (SAE) came up with a definition of car automation levels that is widely used [2, 3, 4, 5]. It is as follows:

A. Level 0 (No Automation): All driving jobs are done by the human operator alone.

B. Level 1 (Driver Assistance): The human driver is in charge of the car, but the automation system helps him or her drive.

C. Level 2 (Partially Automated Driving): The vehicle has a combination of automated functions, but the driver is still in charge of keeping an eye on the surroundings and keeping control of the driving process.

D. Level 3 (Conditional Driving Automation): The human driver must be ready to take charge of the vehicle if needed since the automation system can handle driving in some situations.

E. Level 4 (High Driving Automation): The automation system can drive the car on its own in certain situations, but the human driver may still be able to take over if they want to.

F. Level 5 (Full Driving Automation): The automation system can drive the car on its own in all situations, but the human driver can take over if they want to.

Different groups' definitions of automation levels show that the roles of human drivers and vehicle systems can change in driving. This shows that safety concerns for partly autonomous, highly autonomous, and fully autonomous vehicles can vary a lot. When autonomous cars have different levels of automation, like none, some, or a lot, it's hard to make sure they're safe because people and machines have to work together. On the other hand, when AEVs are operating in fully autonomous states, the software and hardware must be very reliable. As cars add more self-driving technology, the complexity of the self-driving system grows. This makes it harder to keep the system stable, reliable, and safe. To figure out how safe AEVs are now and how safe they will be in the future, it is important to do theoretical studies of possible AEV mistakes [3, 5].

## 3. CATEGORIES OF ERRORS IN AUTONOMOUS ELECTRIC VEHICLES

As the utilization of autonomous techniques increases, the likelihood of encountering various error types rises. Inadequate handling of these errors can give rise to substantial safety implications. Undertaking a systematic analysis of errors and accidents associated with autonomous electric vehicle (AEV) technology is imperative to gain insights into the current state of AEV safety. It is important to note that the reported incidence of accidents involving AEVs is considerably lower than that of traditional vehicles. However, this discrepancy does not inherently imply that current AEVs are inherently safer than human-controlled vehicles. Since AEV technology is still in its nascent stages of commercialization, and complete autonomous driving capabilities remain distant, conducting additional road tests and developing comprehensive accident databases are necessary to achieve a more comprehensive understanding of safety trends [6].

AEV safety hinges on the dependability of the AEV's architecture, encompassing its hardware and software components. However, the architecture of AEVs is intricately linked to the level of automation, thereby resulting in potential variations in AEV safety profiles at different stages of automation. Furthermore, even within the same level of automation, discrepancies in AEV architecture can be observed across different studies. Figure 1 depicts the overarching architecture and key constituents of AEVs. Typically, an AEV consists of a sensor-based perception system, an algorithm-based decision system, an actuator-based actuation system, and interconnected systems [7, 8]. In an ideal scenario, all these components should operate effectively to ensure AEV safety.

### 3.1 Accidents Directly Caused by AEVs
The occurrence of accidents involving autonomous electric vehicles (AEVs) is intricately connected to the occurrence of errors at different levels of automation. These errors can be systematically classified based on the aforementioned architectural framework [5, 7].

### 3.1.1. Perception Inaccuracy

Collecting data from multiple sensing devices is essential for the perception layer to comprehensively understand the environment and make real-time judgments. The development of autonomous electric vehicles (AEVs) relies heavily on the sophistication, reliability, utility, and complexity of sensor technologies. AEVs utilize various technologies, such as LIDAR sensors, cameras, radars, ultrasonic sensors, touch sensors, and GPS, to perceive and interpret their surroundings. Additional information on different sensor systems can be found in other sources [5, 9]. It is crucial to note that a lack of awareness regarding road conditions, the location and movements of other vehicles, traffic signs, and potential hazards can lead to safety challenges.

Figure 2 illustrates the evolution of sensor technologies used in automotive systems over time [5]. This statistic stems from the previously stated sources. Proprioceptive sensors such as wheel sensors, inertial sensors, and odometry were widely used in vehicle systems in the latter half of the twentieth century to improve vehicle dynamics stability and enable functionalities such as traction control, antilock braking, electronic stability control, antiskid control, and electronic stability programs. In the early twenty-first century, exteroceptive sensors such as sonar, radar, LIDAR, vision sensors, infrared sensors, and GPS became more common. By providing navigation help, parking assistance, adaptive cruise control, lane departure warnings, and night vision capabilities, these sensors sought to improve driver information, alarms, and comfort [10].

electric vehicles (AEVs) are prone to perception errors due to concerns related to their hardware, software, and communication systems [11].

The perception system heavily relies on sensing technology, and faulty sensors or equipment can result in incorrect perception. A sensor failure or degradation can lead to significant misinterpretations, confusion in the decision-making process, and hazardous driving situations. Therefore, ensuring the dependability and fault tolerance of sensor technology is crucial [5]. Additionally, perception errors can occur when software faults deliver inaccurate information to the decision and action levels, potentially resulting in mission failure or safety issues.

As AEVs become more automated, communication errors become increasingly dangerous. These issues can arise from AEVs communicating with the internet, other vehicles on the road, and infrastructure [5, 7]. Communication plays a vital role in today's transportation system [6] by facilitating the coordination of all road users, including cars, pedestrians, cyclists, and construction workers, to ensure road safety, which is crucial for AEVs. Communication methods encompass gestures, facial expressions, and in-car electronics. However, the interpretation of these communications can vary based on cultural norms, context, and individual experiences, posing challenges for AEV technology [5, 6, 8].

### 3.1.2. Decision Inaccuracy

The decision layer plays an essential role in examining the processed data from the perception layer, formulating decisions, and generating the necessary information for the action layer [5].

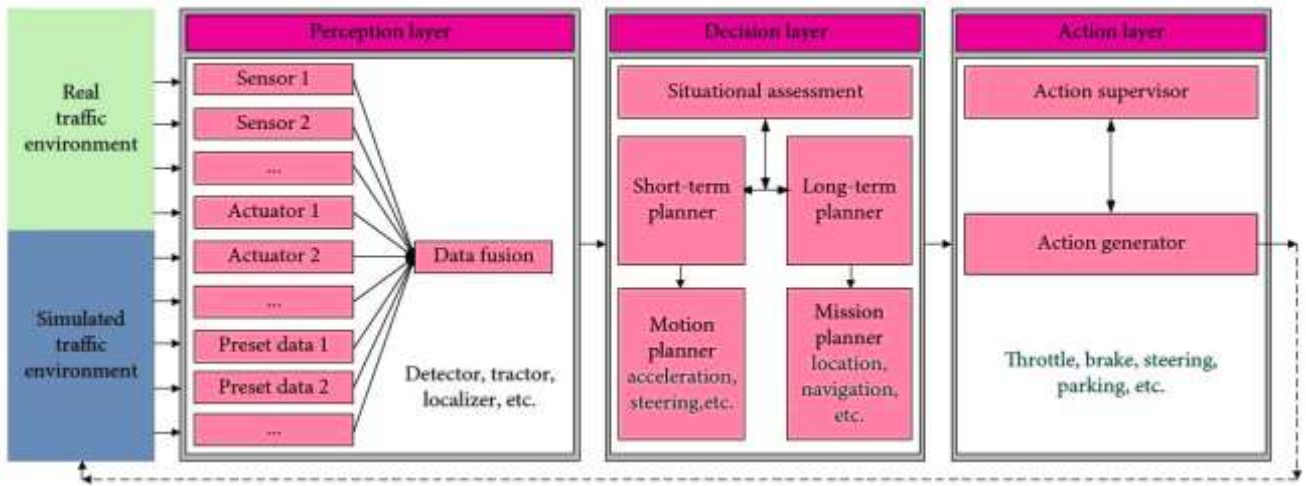

**Figure 1:** Illustrative Architecture for AEV System

In the past decade, sensor networks have been integrated into both roadways and vehicles within modern transportation systems, enabling automatic and collaborative driving [5]. This breakthrough paves the way for advanced autonomous capabilities like collision avoidance and minimization. Ultimately, it leads to fully automated driving, eliminating the need for human drivers. Perceived data can also be obtained through interactions with AEVs, associated infrastructure, other vehicles, the internet, and cloud platforms, depending on the level of vehicle automation. Autonomous

Situational awareness acts as an input for the decision-making system, facilitating both short-term and long-term planning. Short-term planning entails tasks such as generating paths, evading obstacles, and managing incidents and maneuvers, while long-term planning encompasses mission and route planning [12].

Inaccuracies in decision-making primarily arise from factors linked to the system or human involvement. A competent AEV system should intervene or notify the driver only when necessary, upholding a minimal false alarm rate while ensuring acceptable safety performance. With advancements in AEV technology, the false alarm rate can be significantly decreased, maintaining accuracy levels that fulfill safety requirements. However, if the algorithm fails to detect all risks effectively and efficiently, it may jeopardize the safety of AEVs. It is noteworthy that when drivers are engrossed in secondary tasks, there might be a brief delay before they can respond and regain control of the automated vehicle, introducing uncertainties to the secure control of AEVs. Unfortunately, AEV technology is not yet entirely dependable, necessitating human drivers to assume control of the driving process and oversee and monitor the driving tasks when the AEV system fails or its performance is restricted. Nevertheless, this transition in the role of human drivers in AEV driving can result in inattentiveness, diminished situational awareness, and a deterioration in manual driving abilities [5, 7, 13]. Accordingly, the design of AEVs with a human-centered perspective should address the safe and effective re-engagement of the driver when autonomous systems encounter failures.

## 3.2 Accidents Due to Other Road Users

Based on the reported incidents involving autonomous electric vehicles (AEVs) by the United States Department of Motor Vehicles [2, 5, 7], the majority of these incidents are ascribed to other entities on public roadways. These entities, such as motor vehicles, cyclists, and pedestrians (some of whom may be agitated or under the influence), frequently exhibit anomalous behavior that presents challenges even for human drivers. It is imperative to thoroughly investigate how advanced AEVs will react in these perilous scenarios, and it is anticipated that this technology will substantially diminish fatal accidents on roadways. However, autonomous technology is not yet fully matured to cope with highly intricate situations until specific pivotal concerns are resolved. These concerns encompass the effective identification and anticipation of hazardous behaviors stemming from other road users, as well as the accurate decision-making by the autonomous system. The proficient detection of hazards caused by other road users is pivotal for AEVs to actively make determinations and avert potential accidents. AEVs must ascertain whether they must undertake actions that might deviate from traffic regulations to prevent severe or injurious accidents.

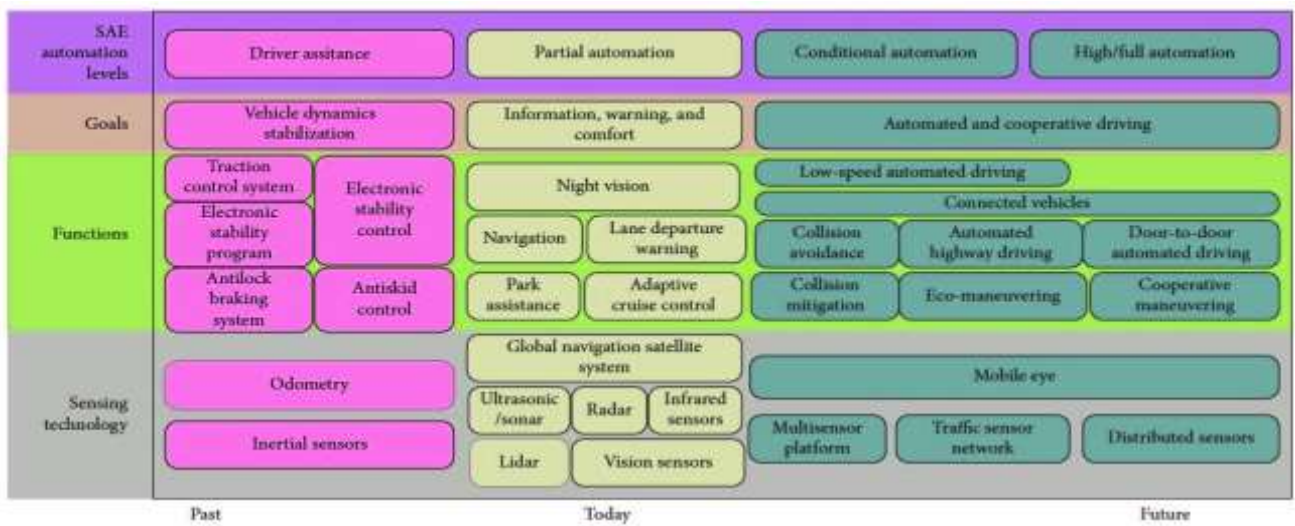

**Figure 2:** Past & Future Development of AEV Technology

### 3.1.3. Action Inaccuracy

Upon receiving instructions from the decision layer, the action controller undertakes further control of the steering wheel, throttle, or brake in the case of a conventional engine [14]. This control enables changes in direction, acceleration, or deceleration. The actuators also monitor feedback variables, utilizing this information to generate new decisions for actuation. Similar to conventional driving systems, inaccuracies in action can arise from actuator failure or malfunctions in the powertrain, control system, heat management system, or exhaust system. These inaccuracies can pose safety risks. However, a human driver is capable of recognizing such safety issues while driving and responding promptly by pulling over.

The challenge for a fully automated driving system lies in how the vehicle learns and responds to these infrequent yet critical malfunctions in major vehicular components. Consequently, the reconstruction of accidents involving traditional vehicles is also of significance [5].

## 4. TESTING AND REPORTING ACCIDENTS (ON-ROAD ANALYSIS)

This section focuses on analyzing publicly available data related to AEV testing, particularly disengagement and accident reports. The objective is to directly assess the safety performance of AEVs. The section explores two key data sources: the California Department of Motor Vehicles (USA) and the Beijing Innovation Center for Mobility Intelligent (China) [5].

### 4.1 DMV – State of California

On-road testing safety issues, like disengagements and actual-life incidents, have been documented by the state's Division of Motor Vehicles [5, 8]. This section focuses on the department's disengagement and collision reports up to April 2019, with an analysis of 621 disengagement reports from 2014 to 2018. Figure 3 displays the cumulative mileage as well as the mileage and disengagement split in California on-road AEV testing, as provided by the Department of Motor Vehicles.

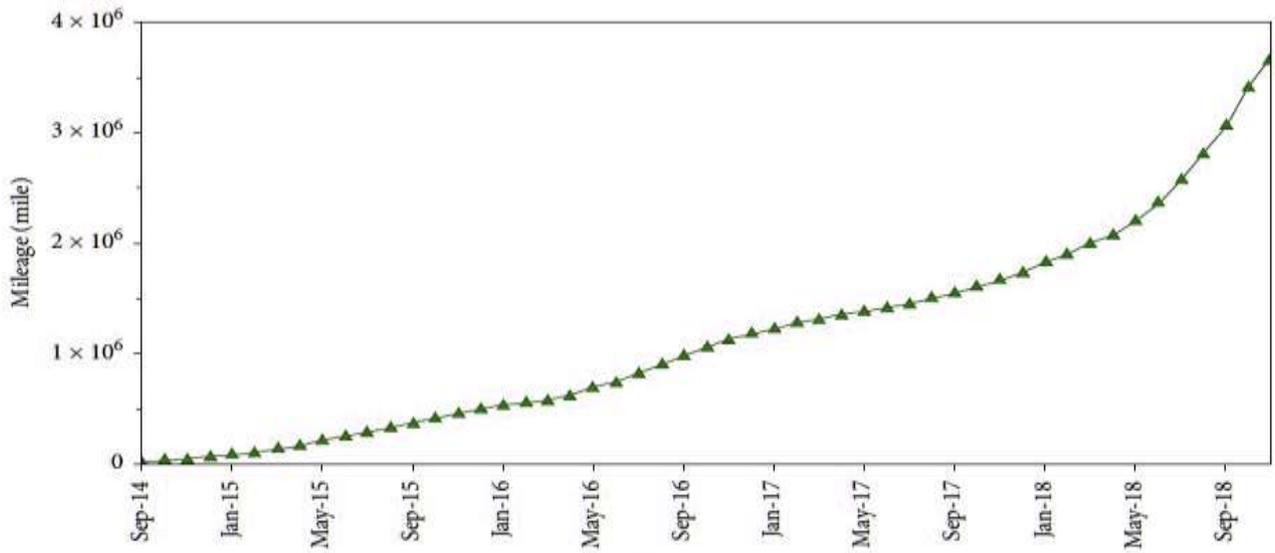

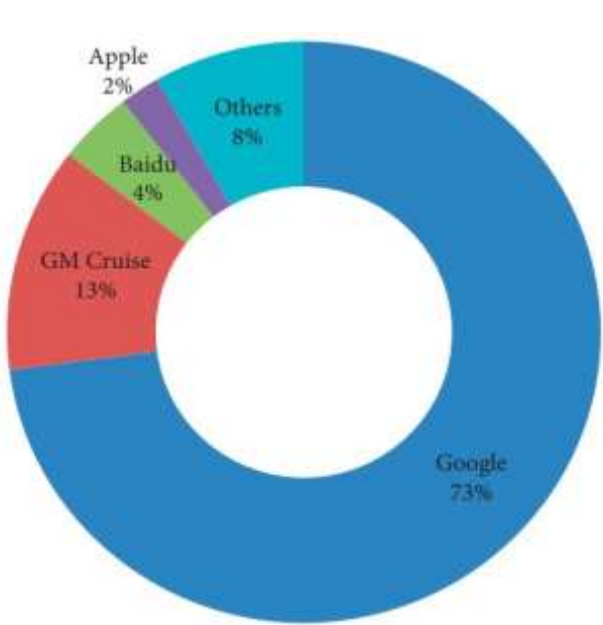
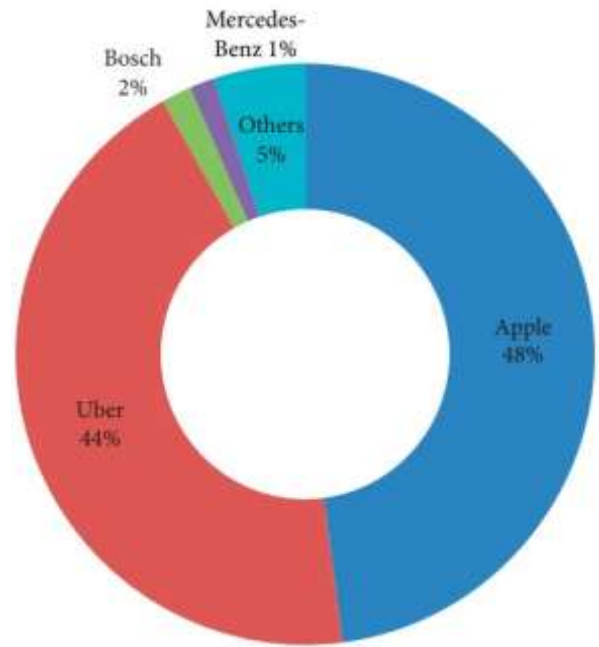

**Figure 3:** (a) Accumulated Distance Traveled, (b) Distribution of Mileage, and (c) Breakdown of Disengagements Across Different Manufacturers

According to a study analyzing 621 disengagement reports (refer to Figure 3(a)), autonomous electric vehicles (AEVs) in California have collectively traveled 3.7 million miles. Google leads the manufacturers in terms of autonomous driving mileage with 73%, followed by GM Cruise at 13%, Baidu at 4%, Apple at 2%, and other manufacturers at 8% (see Figure 3(b)). Apple at 2%, and other manufacturers at 8% (see Figure 3(b)). A total of 159,870 instances of disengagement were documented, with Apple accounting for 48%, Uber accounting for 44%, Bosch accounting for 2%, and Mercedes-Benz accounting for 1%. Disengagement events are classified by Apple into two types: software disengagements and manual takeovers [5].

Instead of depending only on automatic systems, AEV operators have the option of taking manual control of the vehicles when necessary [5]. Figure 3: Statistical analysis was performed on data from the California Department of Motor Vehicles between September 2014 and November 2018; data from Waymo and Google have been pooled and labeled as Google in this figure. These incidents can occur as a result of difficult driving conditions, such as emergency vehicles, construction zones, or unexpected objects on the road. Disengagements in software, on the other hand, are caused by issues recognized in perception, motion planning, controls, and communications.

If the sensors, for example, are unable to detect and track an object in the immediate vicinity, human drivers must take over control of the car. Disengagement events can also occur as a result of the decision layer's inability to generate a motion plan, or as a result of the actuator's delays and incorrect responses. It is important to keep in mind that different manufacturers may interpret disengagement events differently, which implies that reported disengagement events for some organizations may not be full [5, 8]. Figure 4 shows the link between the number of disengagements per mile and the total number of miles for different makers. Manual takeovers happen anywhere from $2 \times 10^{-4}$ to 3 times per mile, depending on the maker. This big difference is mostly caused by differences in the amount of development of autonomous technology. But it's also possible that the way disengagements are described at this early stage of on-road testing leads to differences in how often they happen [5]. Regulators can develop terminologies for disengagement events that take into aspects like perception errors, judgment errors, action errors, system flaws, and other critical factors. These definitions will be extensively distributed.

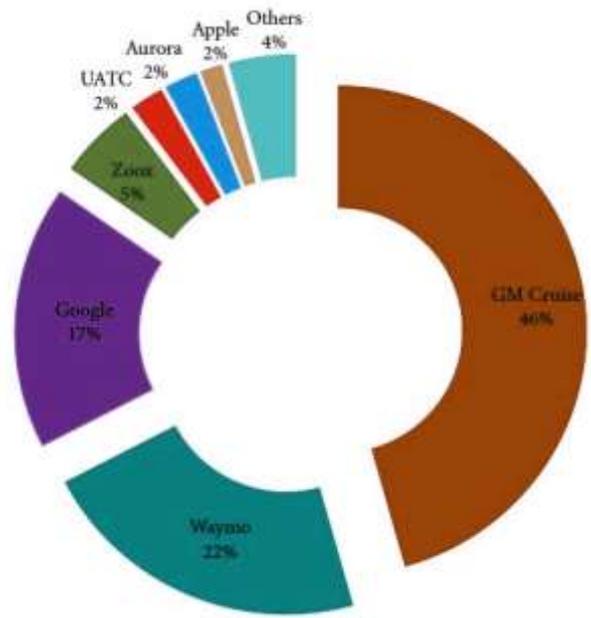

**Figure 5:** Distribution of AEV Accident Reports

happened when the car was being driven by a person instead of by themselves. This shows that driverless technology in AEVs needs to be tested more thoroughly on the road before it can be used everywhere. It's interesting to note that most accidents (93.7% of them) were caused by third parties like walkers, cyclists, motorcyclists, and regular cars, while only a small number (about 6.3%) were caused by the AEVs themselves [5, 15,].

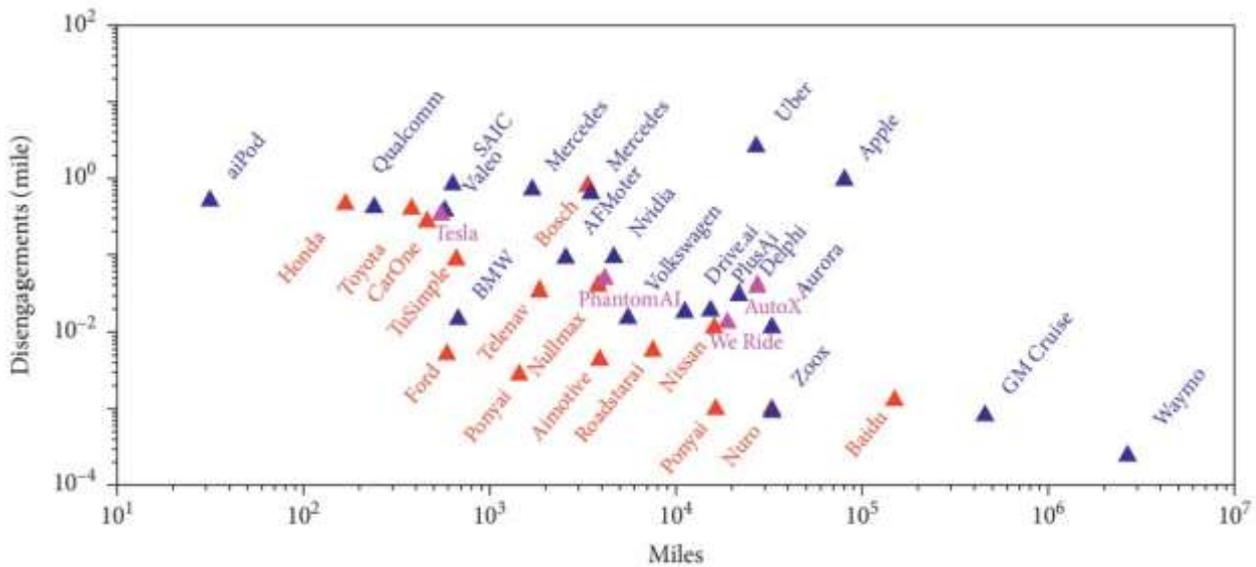

**Figure 4:** Disengagement occurrences plotted against autonomous miles based on reported data provided by different manufacturers

Based on Department of Motor Vehicles information, figure 5 depicts the breakdown of AEV crashes in California from 2014 to 2018. Due to a statistical analysis of 128 collision reports, 46% were triggered by GM Cruise, 22% by Waymo, 17% by Google, and 5% by Zoox. Waymo began in 2009 as the Google Self-Driving Car Project. Most of the 128 crashes that were reported during this time, or 63.3%

This shows how important it is to study how to operate AEVs in the future to cut down on passive crashes and make safety much better. Figure 6 shows the connection between reported events and the total number of AEVs tested in California. Before 2017, the number of crashes that could be reported went up by $1.7 \times 10^{-5}$ per mile.

This was based on the total testing mileage. However, from 2017 to 2018, this rate tripled to $4.9 \times 10^{-5}$ accidents per mile. This change can be attributed to the utilization of advanced, albeit still developing, technology in recent AEV tests, as well as the growing number of concurrently tested AEVs in California. The data presented in this figure are reported by the manufacturers as of April 2019.

## 4.2 Mobility Innovation Center for Intelligent (Beijing)

In 2018, the Beijing Invention Institute for Intelligent Mobility released an analysis on the evaluation of AEVs in urban areas with restricted space and focused populations [27]. By the end of December 2018, self-driving cars had covered a total distance of 153,565 kilometers, equivalent to 95,420 miles (refer to Figure 7(a)). Baidu comprised 90.8% of the manufacturers examined, ahead of Pony.ai (6.6%), NIO (2.7%), and Daimler AG (0.6%) [5, 16]. There have been no instances of disengagement or incidents as of yet. But it would be very helpful for people to have access to information about accidents. This openness could help all automakers get people to buy cars with automatic technology and give customers more faith in AEVs.

## 5. CHALLENGES & OPPORTUNITIES

The progress of AEV technology brings forth a multitude of advantages, such as improving transportation safety, reducing traffic congestion, liberating humans from driving responsibilities, and generating positive economic and environmental effects [4, 5]. Consequently, there is a rising interest in advanced AEV technology within academic and industrial spheres, offering diverse prospects for AEV advancement. Nevertheless, the extensive implementation of AEVs requires substantial experimental efforts to address challenges associated with software, hardware, vehicle systems, infrastructure, and interactions with other road users.

### *5.1 CHALLENGES*

A big problem for AEVs to become widely used is that people are worried about their safety. To get more people to use AEVs, it's important to deal with the following problems [5]:

### *5.1.1. Reducing Perception Inaccuracy*

Inaccuracies in perception make it hard to find, locate, and classify things in the surroundings. Also, making sure AEVs are safe depends on how well they can see and understand human actions like posture, voice, and movement [17].

### *5.1.2. Reducing Decision Inaccuracy*

Creating a method for making decisions that is reliable, strong, and efficient is important if you want to respond to your environment quickly and accurately [18]. To do this, you need to test your hardware and apps thoroughly and carefully. Also, it is still hard to figure out the right thing to do in complicated situations. For instance, when faced with the dilemma of choosing between causing harm to pedestrians or preventing fatal accidents resulting from sudden system failures or mechanical breakdowns, making decisions becomes exceptionally challenging.

### *5.1.3. Reducing Action Inaccuracy*

Establishing a dependable and stable communication link between the actuators and decision systems is essential to ensure the safety of autonomous electric vehicles (AEVs). This allows the actuators to accurately receive and execute commands from both human operators and automated systems, contributing to the overall safety and efficiency of AEVs.

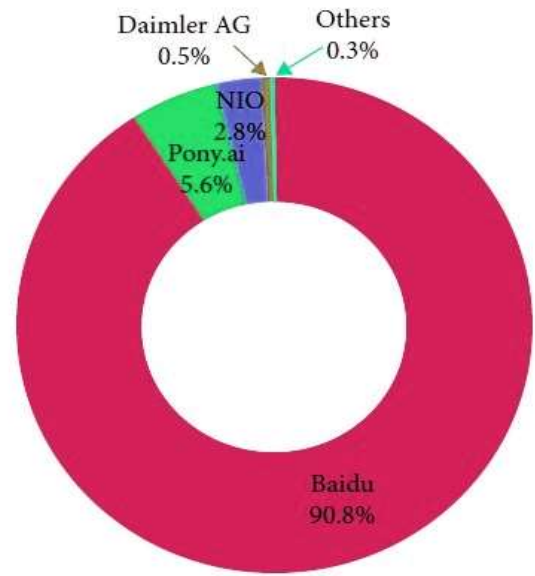

**Figure 6:** Relationship between Cumulative Accidents and Cumulative Autonomous Miles

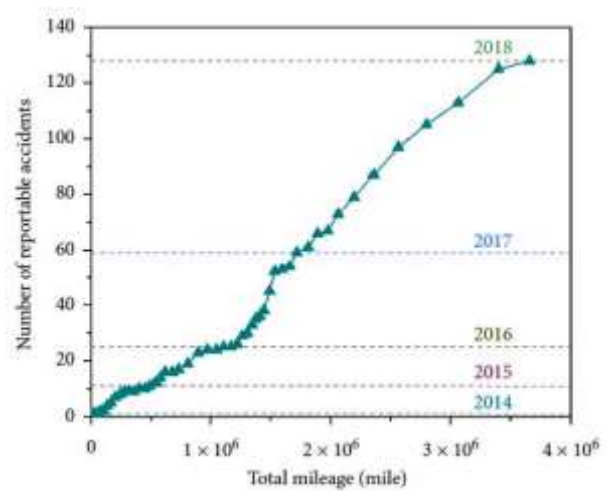

**Figure 7(a):** Cumulative Distance

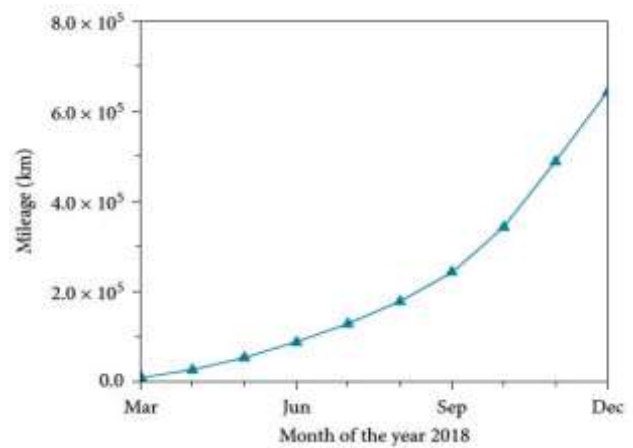

**Figure 7(b):** Comprehensive Analysis of the Distribution of Mileage Contributions among Different Manufacturers

### 5.1.4. Cyber-Security Attack

As autonomous electric vehicles (AEVs) continue to advance, their reliance on wireless connectivity will increasingly extend to interactions with road infrastructure, satellites, and other vehicles, forming what is often referred to as a "vehicular cloud." This is because autonomous technology is getting better. It is of utmost importance to prioritize the implementation of strong cybersecurity measures to address one of the primary concerns related to AEVs [5, 8].

### 5.1.5. Communication with Regular Transportation System

When AEVs and regular cars share public roads in cities, it can be hard for them to get along with other road users, like walkers and drivers of regular cars [18]. It becomes hard for these road users to tell what kinds of vehicles they are dealing with. The ambiguity surrounding AEVs can create feelings of stress for pedestrians and influence their decision-making, particularly when AEV drivers are preoccupied and fail to establish eye contact [5]. Rodrguez Palmeiro et al. propose the utilization of comprehensive behavioral assessments, such as eye-tracking, to gain further insights into pedestrians' reactions toward AEV technology [4].

### 5.1.6. Customer Acceptance

The broad implementation of AEVs encounters notable challenges, including safety considerations, cost implications, and public concerns [8, 9, 10]. Among these factors, safety emerges as the most crucial aspect due to its significant influence on public perception and acceptance of the emerging AEV technology.

### 5.2 OPPORTUNITIES

One argument supporting the development of AEV technology is that although some traditional job opportunities may be eliminated, the overall impact will result in the creation of more jobs. Extensive testing in diverse domains, including software, hardware, vehicle components, vehicle systems, sensing devices, and communication systems, is essential to drive the progress of AEVs [19, 20, 21]. By implementing AEV technology, human operators can be liberated from the driving process, leading to improved time management and increased efficiency in various aspects of life, including work, leisure activities, and education. Furthermore, the adoption of AEV technology brings about lifestyle changes, affecting areas like driving training and driver's license testing. This not only fosters progress within the AEV-related industry but also extends its benefits to non-AEV sectors [22, 23, 24].

AEV techniques can change the way people usually get around. The need for drivers to not have to drive has created an intelligent vehicle grid [25]. The foundation of this system is built upon sensor platforms that gather data from the surrounding environment, including information from other drivers and road signs [26]. These signals are then transmitted to drivers and infrastructure to assist in ensuring safe navigation, reducing pollution, improving gas mileage, and enhancing traffic control [27, 28]. A study conducted by Stern et al. involved an experiment on a ring road, where both autonomous and human-driven cars were present [29]. The results demonstrated that a single AEV could effectively regulate the traffic flow of at least 20 human-driven cars, leading to significant improvements in vehicle speed standard deviation, excessive braking, and fuel economy [30]. Liu and Song looked into two kinds of lanes for AEVs: lanes just for AEVs and lanes for AEVs that charge a fee [31]. Only autonomous cars can use dedicated AEV lanes, but human-driven vehicles can use AEV/toll lanes if they pay extra fees. Their models show that using both types of lanes can make the system work better. Gerla et al. looked at the Internet of Vehicles, which includes the ability to communicate, store information, be smart, and learn on its own [5, 10]. Their work suggests that the communication between vehicles and the Internet will substantially transform public transportation, making it more efficient and environmentally friendly. Consequently, traditional transportation systems must be adapted to accommodate AEVs [5].

Driving simulators have gotten a lot of attention because they can simulate automatic driving and accidents in virtual reality settings. Using driving simulators, researchers can learn a lot about how people drive, requests to take over, car-following moves, and other human factors [5]. This method reduces the risks of putting drivers in dangerous situations while giving them the chance to look at how decisions are made and what happens as a result.

## 6. CONCLUSION

Fully autonomous electric vehicles (AEVs) enable operators to engage in non-driving tasks, providing benefits to individuals and communities. However, the successful commercialization of AEVs faces significant technical challenges due to safety concerns. This comprehensive review article compares automation levels defined by various organizations, with widespread adoption of the Society of Automotive Engineers (SAE) standards. The article also conducts a theoretical analysis of accident types based on typical AEV architectures, encompassing perception, decision, and action systems. Statistical analysis of publicly available on-road AEV disengagement and accident reports in California reveals over 3.7 million miles of testing conducted by different manufacturers between 2014 and 2018. Disengagement frequencies range from $2 \times 10^{-4}$ to 3 disengagements per mile, varying among manufacturers. Among the 128 reported accidents, approximately 63.3% occurred during the autonomous mode, with only around 6.7% directly attributed to AEVs, while 94.7% were caused by pedestrians, cyclists, motorcycles, and conventional vehicles [5, 7]. These findings emphasize the need to address safety risks posed by other road users and make informed decisions to prevent fatal accidents.

## Conflicts of Interest

The authors declare that they have no conflicts of interest.

*Correspondences should be addressed to Qasim Ajao; Qasim.ajao@ieee.org